\documentstyle{article}

\begin{document}

\begin{center}
\em Published in {\bf Fizika} (Zagreb) {\bf 21}, suppl.\ 3, 231 (1989)

Proceedings of the Third European Conference on Low Dimensional
Conductors and Superconductors

Dubrovnik, September 18--22, 1989

\bf cond-mat/9703195
\end{center}

\begin{center}
{\bf Spin, Statistics and Charge of Solitons in (2+1)-Dimensional
Theories}\\
\bigskip
Victor M. YAKOVENKO\\
\bigskip
L.D. Landau Institute for Theoretical Physics,
2 Kosygin St., Moscow 117940, USSR \\
\end{center}

{\bf Abstract.} General topologically invariant microscopical
expressions for quantum numbers of particle-like solitons
(``skyrmions'') are derived for a class of (2+1)D models.  Skyrmions
are either half-integer spin fermions with odd electric charge or
integer spin bosons with even charge. So they cannot be Anderson's
spinons or holons. General results are exemplified by a square lattice
model reminiscenting high-$T_c$ models.
\bigskip

It was conjectured by P.W.Anderson \cite{A} that unusual particles ---
spinons and holons --- may exist in condensed matter.  Spinons are
neutral fermions with spin $\hbar/2$ while holons are spinless bosons
with charge $e$. In polyacetylene, which is (1+1)D system, solitons
really have such quantum numbers \cite{B}. In (2+1)D case it is
interesting to consider \cite{DPW} quantum numbers of ``skyrmions''
--- particle-like solitons of $\vec{n}$-field. To describe skyrmion
let us map $(x,y)$ plane on the sphere $S^2$ by stereographic
projection. The skyrmion configuration is the hedgehog configuration
of $\vec{n}$-field on the sphere $S^2$. At the $(x,y)$ plane this
configuration looks as follows: $\vec{n}$ is down at the center,
$\vec{n}$ is up at the infinity and there is a concentric domain wall
in between.

Quantum numbers of skyrmions can be calculated in a general form
\cite{VY}, \cite{Y} for a class of models which have the action
\begin{eqnarray}
    &S(\psi(\vec{r}),\vec{n}(\vec{r}))=Tr\int d^3r_1 d^3r_2
    \psi^+(\vec{r_1})\tilde{G}^{-1}(\vec{r_1},\vec{r_2})
    \psi(\vec{r_2}),&
\label{S}\\
    &\tilde{G}^{-1}(\vec{r}_1,\vec{r}_2)
    =G^{-1}_0(\vec{r}_1,\vec{r}_2)
    +\vec{\sigma}\left[\vec{n}(\vec{r}_1)+\vec{n}(\vec{r}_2)\right]
    G^{-1}_1(\vec{r}_1,\vec{r}_2).&
\label{G01}
\end{eqnarray}
Here $\vec{r}_1,\vec{r}_2$ are (2+1)D space-time coordinates
(continuous or discrete), $\psi(\vec{r})$ is an electron annihilation
operator, $\tilde{G}^{-1}$ is the inverse Green function taken in the
real-space representation, $\vec{\sigma}$ are Pauli matrices which act
on the spin indices of electrons.  Matrix functions $G^{-1}_0$ and
$G^{-1}_1$ are the coefficients of decomposition of the inverse Green
function $\tilde{G}^{-1}$ over the unit and the Pauli spin
matrices. Functions $G^{-1}_0$ and $G^{-1}_1$ have translational
symmetry appropriate to uniform or periodic potential, while
$\vec{n}(\vec{r})$ is any vector function slowly varying in space-time
with the constrain $|\vec{n}(\vec{r})|=1$.  Trace is taken over spin
and other indices of electrons which are assumed. The system of units
, where $e,\hbar$ and $c$ are set to unity, is used. Dimensional
factors are restored only in final formulas.

Physically action (\ref{S}), (\ref{G01}) describe electrons which do
not interact between each other but interact with external $\vec{n}$-
field.  So (\ref{S}), (\ref{G01}) is typical mean-field action which
frequently appears in weak-coupling models. Some people try to
describe even strongly correlated electrons by effective action of
this type.

Now let us perform functional integration over $\psi(\vec{r})$ field
and find effective action $S_{eff}(\vec{n}(\vec{r}))$ for
$\vec{n}$-field.  This action contain the term
\begin{equation}
    S_{1}=\frac{C_1\varepsilon_{\mu\nu\lambda}}{32\pi}
    \int d^3r\;B_\mu P_{\nu\lambda},\;\;\;
    P_{\mu\nu}=\vec{n}\left[
    \frac{\partial\vec{n}}{\partial r_\mu}\times
    \frac{\partial\vec{n}}{\partial r_\nu}\right],\;\;\;
     \frac{\partial B_\nu}{\partial r_\mu}
    -\frac{\partial B_\mu}{\partial r_\nu}=P_{\mu\nu}.
\label{S1}
\end{equation}
The coefficient $C_1$ is given \cite{VY} by the following expression
\begin{equation}
    C_1=N(G),\;\;\;
    N(G)=\frac{\varepsilon_{\mu\nu\lambda}}{24\pi^2}
    Tr\int d^3k\;G\frac{\partial G^{-1}}{\partial k_\mu}
    G\frac{\partial G^{-1}}{\partial k_\nu}
    G\frac{\partial G^{-1}}{\partial k_\lambda}.
\label{NG}
\end{equation}
In formula (\ref{NG}) we use Green function (\ref{G01}) in the
momentum representation $G(\vec{k})$ with $\vec{n}(\vec{r})$ field set
to uniform distribution $\vec{n}(\vec{r})=\vec{n}_0=\vec{e}_z$ where
$\vec{e}_z$ is a unit vector along $z$-axis in spin space. Integral
(\ref{NG}), taken over momenta and frequency, is a topological
invariant of the matrix function $G(\vec{k})$. It takes integer
values. As was shown in \cite{WZ} skyrmions have spin
\begin{equation}
    S=C_1\hbar/2
\end{equation}
and corresponding statistics. Thus they may have either fermion or
boson but not fractional statistics depending on the parity of integer
(\ref{NG}). Taking into account that $G(\vec{k})$ is diagonal in spin
indices expression (\ref{NG}) can be rewritten as
\begin{equation}
    C_1=N(G_\uparrow)+N(G_\downarrow),
\label{N+N}
\end{equation}
where
$   G^{-1}_{\uparrow,\downarrow}(\vec{k})
    =G^{-1}_0(\vec{k})\pm 2G^{-1}_1(\vec{k})$.

To find electrical charge of the skyrmion the electromagnetic
potential $A_\mu(\vec{r})$ is introduced in (\ref{S}) in standard
way. After functional integration over fermions two additional to
(\ref{S1}) terms appear in $S_{eff}(\vec{n}(\vec{r}),A_\mu(\vec{r}))$:
\begin{equation}
    S_{2}=\frac{C_2\varepsilon_{\mu\nu\lambda}}{4\pi}
    \int d^3r\;A_\mu
    \frac{\partial B_\lambda}{\partial r_\nu},\;\;\;
    S_{3}=\frac{C_3\varepsilon_{\mu\nu\lambda}}{4\pi}
    \int d^3r\;A_\mu
    \frac{\partial A_\lambda}{\partial r_\nu}.
\label{S3}
\end{equation}
Varying $S_2$ over $A_0$ we find expression for the electric charge of
the skyrmion:
\begin{equation}
    e^*=C_2e.
\end{equation}
Varying $S_3$ over $A_x$ we find nondiagonal component of conductivity
\begin{equation}
    \sigma_{xy}=C_3e^2/h.
\end{equation}
Microscopical expressions for $C_2$ and $C_3$ are as follows \cite{VY},
\cite{Y}, \cite{V}:
\begin{equation}
    C_2=N(G_\uparrow)-N(G_\downarrow),\;\;\;
    C_3=C_1=N(G_\uparrow)+N(G_\downarrow).
\label{C23}
\end{equation}
From (\ref{C23}) we see that fermions always have odd charge while
bosons have even charge. So skyrmions cannot be spinons or holons.

Now let us consider a particular model where $C_1,C_2$ and $C_3$ have
non-zero values. It is square lattice model with the period doubling
in both directions. Green functions
$G_\alpha(\vec{k}),\;\;\alpha=\uparrow,\downarrow$ can be expanded
over Pauli matrices $\vec{\tau}$ which act on the doublet of electrons
$(\psi(k_x,k_y),\psi(k_x+\pi,k_y+\pi))$:
\begin{eqnarray}
    &G_\alpha(\vec{k})=i\omega+\vec{\tau}\vec{w}(k_x,k_y),&
\label{w}\\
    &w_1=M_\alpha+t_2(\cos(k_x+k_y)-\cos(k_x-k_y)),&\\
    &w_2=J(\cos k_x-\cos k_y),\;\;\;w_3=t_1(\cos k_x+\cos k_y).&
\label{W}
\end{eqnarray}
In (\ref{W}) the term proportional to $t_1$ describes electron hopping
between the nearest neighboring sites. The term proportional to $J$
describes current density wave that is spontaneous staggering currents
along the links of square lattice. It was suggested in \cite{HR} and
recently discussed in \cite{NL}. With $J=t_1$ and $t_2=M=0$ model
(\ref{w})--(\ref{W}) is very close to the ``flux phase'' model
\cite{AM}. But in weak coupling limit relation between $J$ and $t_1$
is arbitrary. The term proportional to $t_2$ was suggested in
\cite{WWZ} and describe staggering hopping amplitude between the
next-nearest neighboring sites. This term breaks macroscopic
time-reversal and inversion symmetries of the model. The terms
proportional to $M_\alpha$, suggested in \cite{Sch}, describe
staggering site energy.  There are charge density wave with the
amplitude $(M_\uparrow+M_\downarrow)/2$ and spin density wave with the
amplitude $(M_\uparrow-M_\downarrow)/2$ in the model.  Polarization of
SDW is determined by $\vec{n}$-field, which is set to the constant
value $\vec{n_0}=\vec{e_z}$ in (\ref{w})--(\ref{W}).

Applying the methods of \cite{VY}, \cite{Y} and \cite{V} to model
(\ref{w})--(\ref{W}) we find: $N(G_\alpha)$=sign($t_2$) if
$|2t_2|>|M_\alpha|$ and $N(G_\alpha)=0$ otherwise.  Let
$|M_\uparrow|>|M_\downarrow|$. Then from (\ref{C23}) it follows that
in the region $|2t_2|>|M_\uparrow|$ skyrmions are neutral bosons with
spin $\hbar$, in the region $|M_\downarrow|<|2t_2|<|M_\uparrow|$
skyrmions are fermions with spin $\hbar/2$ and charge $e$, in the
region $|2t_2|<|M_\downarrow|$ skyrmions are neutral bosons with spin
$0$.  If $M_\alpha=0$ and $J$ has opposite signs for spins up and down
then skyrmions are spinless bosons with charge $2e$. In this case the
staggering currents are the spin currents and their polarization is
given by $\vec{n}$-field.


\begin{thebibliography}{12}
\parskip    0pt
\itemsep    0pt

\bibitem{A}  P.W.Anderson, ``50 years of Mott phemonena'', Varenna,
             July 1987.
\bibitem{B}  S.A.Brazovskii and N.N.Kirova, {\em Soviet Scientific
             Reviews} {\bf A5}, 99 (1984).
\bibitem{DPW}I.Dzyaloshinskii, A.Polyakov and P.Wiegmann,
             {\em Phys. Lett.} {\bf A 127}, 112 (1988).
\bibitem{VY} G.E.Volovik and V.M.Yakovenko,
             {\em J. Phys.} {\bf CM 1}, 5263 (1989).
\bibitem{Y}  V.M.Yakovenko, {\em Phys. Rev. Lett.} {\bf 65}, 251 (1990).
\bibitem{WZ} F.Wilczek and A.Zee, {\em Phys. Rev. Lett.} {\bf51}, 2250
             (1983).
\bibitem{V}  G.E.Volovik {\em Zh. Exp. Teor. Fiz.} {\bf 94}, 123
             (1988).
\bibitem{HR} B.I.Halperin and T.M.Rice, {\em Sol. State Phys.} {\bf 21},
             115 (1968).
\bibitem{NL} A.A.Nersesyan and A.Luther, NORDITA preprint 1988
             (unpublished), H.J.Schulz, {\em Phys. Rev.} {\bf B 39},
             2940 (1989),
             I.E.Dzyaloshinskii and V.M.Yakovenko, {\em Int. J. Mod.
             Phys.} {\bf B 1}, 667 (1988).
\bibitem{AM} I.Affleck and B.J.Marston, {\em Phys. Rev.} {\bf B 37}, 3774
             (1988).
\bibitem{WWZ}X.G.Wen, F.Wilczek and A.Zee, {\em Phys. Rev.} {\bf B
             39}, 11413 (1989).
\bibitem{Sch}H.J.Schulz, private communication.
\end{thebibliography}
\end{document}